\documentstyle[twoside,fleqn,espcrc2,epsf]{article}

\newcommand{\AmS}{{\protect\the\textfont2
  A\kern-.1667em\lower.5ex\hbox{M}\kern-.125emS}}

\title{Finite Lattice Hamiltonian Computations in the
P-Representation: the Schwinger Model}

\author{J.M. Aroca\address{Departament de 
Matem\`atiques, Universitat Polit\`ecnica de Catalunya,\\
Jordi Girona 1-3, Mod C-3 Campus Nord,\\
08034 Barcelona, Spain.}%
\thanks{Supported in part by CIRIT, project ACI014.}
, H. Fort\address{Instituto de F\'{\i}sica, 
Facultad de Ciencias,  
Tristan Narvaja 1674, 11200 Montevideo, Uruguay}%
\thanks{Supported in part by CONICYT, Project No. 318.}
and Gonzalo Alvarez$^{\rm b}$ 
}

\begin{document}
\input psfig

\begin{abstract}

The Schwinger model is studied in a finite lattice
by means of the P-representation. The vacuum
energy, mass gap and chiral condensate are evaluated
showing good agreement with the expected values
in the continuum limit.
\end{abstract}

\maketitle

\section{Introduction}

A useful formulation of gauge theories, both from the 
conceptual and methodological point of view, is the one in terms
of gauge invariant excitations or string-like objects.
The so-called {\em P-representation} \cite{fg}, consisting
of a Hilbert space of path labeled states, has been used on the lattice
to perform analytical Hamiltonian
calculations. A cluster approximation allowed to provide 
qualitatively good results for the $(2+1)$ QED \cite{af2d}
and the $(3+1)$ QED \cite{af3d} with staggered fermions.
A description in terms
of paths or strings, besides the general advantage of
only involving gauge invariant excitations, is appealing because
all the gauge  invariant  operators have a 
simple geometrical meaning when
realized in the path space.  
However, the computational method implemented, up to now, on a formally
{\em infinite} lattice,  has the serious
drawback of the explosive proliferation of 
clusters with the order of the approximation.
In order to tackle this difficulty we propose in this
paper to explore the previous method implemented now on a 
{\em finite lattice}.
As a first test, we choose the simplest lattice 
gauge theory with dynamical
fermions, the Schwinger model or (1+1) QED.
This massless model can be exactly solved in the continuum
and it is rich enough to share relevant features 
with 4-dimensional QCD as confinement or
chiral symmetry breaking with an axial anomaly \cite{col76}.
For this reason it has been extensively used as a laboratory to 
analyze the previous phenomena.
The lattice Schwinger model also become a popular
benchmark to test different techniques to handle
dynamical fermions \cite{hamer80}-\cite{hamer97}.

This article is organized into four sections. In section 2
we show the formulation of the model in the P-representation.
The electric and interaction components of the Hamiltonian 
operator are realized in this basis of ``electromeson" states.
In section 3, first, we describe the finite lattice Hamiltonian 
approach. Second, we show the calculation of 
the ground-state energy, the 
mass gap and chiral condensate.
These results are discussed in the concluding section.  

\section{Schwinger Model in the Lattice P-Representation}

        The P-representation offers a 
gauge invariant description of physical
states in terms of kets $\mid P >$, 
where $P$ labels a set of connected paths
$P_x^y$ with ends $x$ and $y$ 
in a lattice of spacing $a$.  
In the continuum, the connection
between the P-representation and the ordinary 
representation (``configuration" representation), 
in terms of the fermion fields $\psi$ and the gauge 
fields $U_\mu(x)=\exp [iea A_\mu(x)]$, can be performed 
considering the natural gauge invariant object constructed
from them:
\begin{equation}
\Phi (P_x^y) = {\psi}^{\dagger} (x) U(P_x^y)\psi (y),
\end{equation}
where $U(P_x^y)=\exp [ie \prod_{\ell \in P} A_\ell]$
( $\ell \equiv (x, \mu)$ denote the links ).

The immediate problem we face is that $\Phi$ 
is not purely an object belonging
to the ``configuration'' basis because it 
includes the canonical conjugate
momentum of $\psi$, ${\psi}^{\dagger}$.  
The lattice offers a solution to this problem consisting in
the decomposition of the fermionic degrees 
of freedom. Let us consider the
Hilbert space of kets $\mid {\psi}_u^{\dagger} 
,{\psi}_d , A_{\mu} >$, where
$u$ corresponds to the $up$ part of the Dirac 
spinor and $d$ to the $down$
part. Those kets are well defined in terms 
of ``configuration'' variables (the
canonical conjugate momenta of ${\psi}_d$ and  
${\psi}_u^\dagger$ are
${\psi}_d^\dagger$ and ${\psi}_u$ respectively.)
Then, the internal product of
one of such kets with one of the path dependent 
representation (characterized
by a lattice path $P_x^y$ with ends $x$ and $y$) is given by
$$
\Phi (P_x^y) \equiv <P_{x;i}^{y;j} \mid {\psi}_u^{\dagger} ,
{\psi}_d, A_{\mu} > 
$$
\begin{equation}
= {\psi}_{u;i}^{\dagger} (x) U(P_x^y) {\psi}_{d;j}(y),
\label{eq:Phi}
\end{equation}
where $i$ and $j$ denote a component of the 
spinor $u$ and $d$ respectively.
Thus, it seems that the choice 
of staggered fermions is the natural one 
in order to build the lattice P-representation.  
Therefore, the lattice 
paths $P_x^y$ start in
sites $x$ of a given parity and end in sites 
$y$ with  opposite parity.
The one spinor component at each site 
can be described in terms of the Susskind's 
$\chi (x)$ single Grassmann fields \cite{sus}. 
The path creation  
operator $\hat{\Phi}_Q$ in the space of kets $\{ |\, P> \}$
of a path with ends $x$ and $y$ is defined as 
\begin{equation}
\hat{\Phi}_Q= \hat{\chi}^{\dagger} (x) \hat{U}(Q_x^y) 
\hat{\chi}(y).
\label{eq:Phiop}
\end{equation}
Its adjoint operator $\hat{\Phi}_Q^{\dagger}$ acts
in two possible ways \cite{fg}: annhilating the path $Q_x^y$
or joining two existing paths in $|\, P>$ one ending at
$x$ and the other starting at $y$.

The Schwinger Hamiltonian is given by

$$
\hat{H}=\frac{ae^2}{2}\hat{W}
$$
$$
\hat{W}=\hat{W}_E +\lambda \hat{W}_I \, , \;\;\;
\lambda =\frac{1}{a^2e^2}  \, ,
$$
\begin{equation}
\hat{W}_E=\sum_\ell \hat{E}_\ell^2 \, , 
\label{eq:H}
\end{equation} 
$$
\hat{W}_I=\sum_{\ell} (\hat{\Phi}_\ell + 
\hat{\Phi}_\ell^{\dagger}) \hspace{3mm}
(\hat{\Phi}_\ell=\hat{\chi}^{\dagger}(x)
\hat{U}_n(x)\hat{\chi} ( x+ n) )
$$
where  $x$ labels sites, $\ell \equiv (x, n)$ the spatial
links pointing along the spatial unit vector $n$, 
$\hat{E}_\ell$ is the electric field operator, 
the kets $|\, P \, >$ are eigenvectors of this operator
\begin{equation}
\hat{E}_\ell |\, P \, > = N_\ell (P) |\, P \, >,
\label{eq:E}
\end{equation}
where the eigenvalue $N_\ell(P)$ is the number of times that the 
link $\ell$ appears in  the set of paths $P$.
The $\hat{\Phi}_\ell$ are 
``displacement" operators corresponding to the quantity 
defined in (\ref{eq:Phi}) for the case of a one-link
path i.e. $P\equiv \ell$. The realization of both 
Hamiltonian terms in this representation is
as follows \cite{fg}:

\vspace{3mm}

By (\ref{eq:E})
the action of the electric Hamiltonian is given by
\begin{equation}
\hat{W}_E\mid P >=\sum_\ell N_\ell^2(P)\mid P >.
\label{eq:We}
\end{equation}
The interaction term $\hat{W}_I$ 
is realized in the loop space as      
\begin{equation}
\hat{W}_I\mid P >=\sum_{\ell}\epsilon (P,\ell)
\mid P\cdot \ell>
\label{eq:Wi}
\end{equation}
where 
the factor $ \epsilon (P,\ell )$ 
is 0 or $\pm 1$ dictated by the algebra of the operators.
The different actions of operators
$\hat{\Phi}_\ell$ over path-states $|P(t)>$ 
and their corresponding $ \epsilon (P,\ell )$
are schematically summarized in FIG.1.

\begin{center}
\begin{figure}[t]
\hskip 1cm \psfig{figure=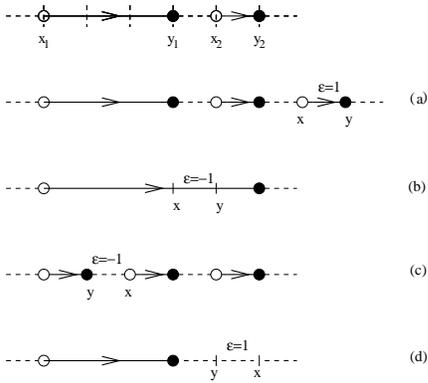,height=5cm}
\vskip -6mm
\caption{A summary of the different actions of
operators $\Phi_\ell$, $\ell$ from $x$ to $y$,
 applied over path-states $|\,P \,>$
with their respective $\epsilon$.
The original path is on the top.
The resulting paths $|P'>$ are plotted below: (a) represent 
the addition of a disconnected link, (b) the union of 2 disconnected
pieces, (c) the separation of a connected piece into 2 pieces
and (d) the annihilation of a one-link path.}
\label{fig1}
\vskip -5mm
\end{figure}
\end{center}

\section{Finite Lattice Hamiltonian Method and Results}

Our method of calculation works assuming a lattice of
some fixed even number of sites $N$ and periodic boundary 
conditions (PBC).  Starting with the zero-path 
state $|\, \emptyset \, >$ (infinite coupling vacuum), then 
a collection of new states
$|\, P_i \, >$ 
are generated by applying successively the non-diagonal $W_I$ 
interaction Hamiltonian operator 
-- whose action is to add or to eliminate links to to the path 
$P_{i-1}$ as it was
described in the previous section --
up to order $K$. 
The traslational symmetry can be exploited 
in order to reduce the dimension of the space 
tacking only one representative 
$\bar{P}_\alpha$ of
each class of translationally equivalent paths $\{ P_i \}$.
The Hamiltonian matrix, with all the 
transitions between the different states
$|\, \bar{P}_\alpha \, >$, is then built  
for the scalar and vector sectors
and their eigenvalues $\omega_i$ are numerically evaluated.

In order to perform 
the generation and recognition of
diagrams (the elementary lattice paths) as well as 
the computation of transitions
between them,
we resorted to the PROLOG language which is very suitable 
to carry out the symbolic manipulations.

\vspace{1mm}

The calculations of the ground-state energy, mass gap and
chiral order parameter were performed on lattices ranging from
size $N=2$ to $N=16$ 
and at least up to order $K=N$ in each case.
Our aim is to extrapolate these results to the continuum
limit: $N\rightarrow \infty$, $a \rightarrow 0$
($\lambda \rightarrow \infty$.)

It is clear from the plots (FIGS. 2 to 5) that the lattice 
results show
convergence to the expected continuum values. This convergence
is, however, non-uniform and for $\lambda$ large enough
the plots show deviation from the continuum values
although the region of assimptotic regime 
becomes larger when the size is increased. It is patent
that for a fixed lattice size $N$ the best results for the vacuum energy
and the chiral condensate are achieved for order $N-2$.
This appears to be the order at which the finite size effects
are minimized. This is not the case with the mass gap which
always gets closer to the continumm value when the order
increases.

\vspace{2mm}

{\em GROUND STATE ENERGY}

\vspace{2mm}

In the continuum limit the ground-state energy density
is known
exactly \cite{hamer80}: 
\begin{equation}
\lim \frac{\omega_0}{2N\lambda}=-\frac{1}{\pi}=-0.3183.
\label{eq:E-exact}
\end{equation}
When the order increases $\omega_0/(2N\lambda)$ tends to a
fixed value. For a fixed size $N$ the closer value to
(\ref{eq:E-exact}) is given by order $N-2$. The value
for size $N=16$ and order $K=14$ at $\lambda =1000$ is 
$\omega_0/(2N\lambda) =0.31844$,  
so the discrepancy from the exact value is less than 0.05 \%.
The approximations converge with considerable rapidity.
FIG. 2 shows $\frac{\omega_0}{2N\lambda}$ 
for orders $K=1,2,\ldots ,14$ for $\lambda$ ranging from 
0 to 100
on a lattice of size $N=14$.

\begin{center}
\begin{figure}[t]
\psfig{figure=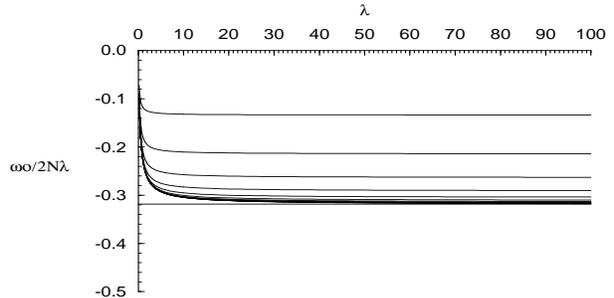 ,height=4.5cm ,width=8.6cm}
\vskip -13mm
\caption{The ground-state energy density 
over $2\lambda^{1/2}$ for orders $K=1,2,\ldots ,14$
on a lattice of size $N=14$. }
\label{fig2}
\end{figure}
\end{center}
\vskip -10mm

In order to obtain a result in a consistent way we compute
the energy
for two large values of $\lambda$, for three correlative
large orders and for three correlative large sizes.
Then, for fixed size and order we first extrapolate 
to $\lambda =\infty$ assuming
the behaviour $a+b/\lambda$. Second, for fixed size
we extrapolate to infinite order
assuming exponential dependence. Finally we extrapolate to
infinite size assuming exponential behaviour. The results
are given in TABLE 1. 
The error using lattice sizes up to $N=16$ is 0.17\% .


\begin{table*}[hbt]
\setlength{\tabcolsep}{1.5pc}
\newlength{\digitwidth} \settowidth{\digitwidth}{\rm 0}
\catcode`?=\active \def?{\kern\digitwidth}
\caption{Ground state energy 
for different lattice sizes and orders 
at $\lambda =\infty$.}
\label{tab:gse}
\begin{tabular*}{\textwidth}{@{}l@{\extracolsep{\fill}}rrrr}
\hline
                 & \multicolumn{1}{r}{$N=12$} 
                 & \multicolumn{1}{r}{$N=14$} 
                 & \multicolumn{1}{r}{$N=16$} 
                 & \multicolumn{1}{r}{$N=\infty$}     \\
\hline
  $K=N-4$   & $-0.317075$ & $-0.317723$ & $-0.318023$ & \\
\hline  
 $K=N-3$    & $-0.318010$ & $-0.318249$ & $-0.318348$ & \\
\hline
 $K=N-2$    & $-0.318697$ & $-0.318656$ & $-0.318608$ &  \\
\hline
 $K=\infty$ & $-0.320611$ & $-0.320045$ & $-0.319660$ & $-0.318847$ \\
\hline
\multicolumn{5}{@{}p{120mm}}
{}
\end{tabular*}
\end{table*}

\vspace{2mm}

{\em MASS GAP}

\vspace{2mm}

The mass gap for the massless continuum Schwinger model
can be computed exactly \cite{s62}:
\begin{equation}
\frac{M^c}{e^c}=\frac{1}{\pi^{1/2}}=0.564,
\label{eq:mg_c}
\end{equation}

The lattice mass gap is computed as:
\begin{equation}
\frac{M}{e}=\frac{\omega_1 -\omega_0}{2\sqrt{\lambda}}
\end{equation}
Comparing our results with those
of Crewther and Hamer \cite{hamer80} obtained by a similar 
method, although they use a different representation (Jordan-Wigner
transformation), we find complete agreement for given values
of $N$ and $K$. When we reach larger $N$ we observe that the value
of the mass gap improves substantially. For instance, in FIG. 3 
we show a plot of the mass gap for $N=10$ for several orders.
As it can be seen in the region $10<\lambda < 30$
, the mass gap values decrease
with the order and the size approaching the continuum result.
Given the non-uniformity of the convergence it is more
difficult to extrapolate to the limit although values
$\approx 0.579$ are obtained at the modest size of $N=8$.





\begin{center}
\begin{figure}[t]
\hskip 1cm \psfig{figure=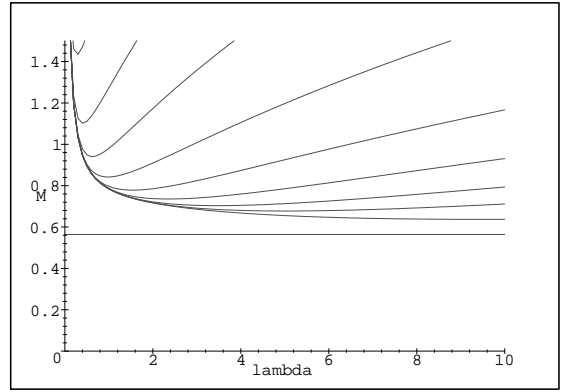,width=8.6cm}
\caption{The mass gap over $2\lambda^{1/2}$ vs. $\lambda$
for orders $K=1,2,..,10$
on a lattice of size $N=10$ .}
\label{fig3}
\end{figure}
\end{center}

\vspace{3mm}

{\em CHIRAL ORDER PARAMETER}

\vspace{2mm}

An interesting quantity to compute is the vacuum expectation
of the chiral condensate per-lattice-site
$<\bar{\chi}\chi >$, defined as
\begin{equation}
\bar{\chi}\chi =\frac{1}{2N_s}\sum_x (-1)^{x}
[\hat{\chi}^\dagger 
(\mbox{\bf x}),\hat{\chi}(\mbox{\bf x})],
\label{eq:chir-cond}
\end{equation}
where $N_s$ is the number of lattice sites.
The corresponding operator is realized in the 
P-representation and thus we get for 
the chiral condensate:
\begin{equation}
\bar{\chi}\chi |P>=
(\frac{1}{2}-\frac{2{\cal N}_P}{N_s})|P>,
\label{eq:chir-cond2}
\end{equation}
where ${\cal N}_P$ is the number of connected paths in $P$.
To compute 
$<\bar{\chi}\chi >$ the $\hat{W}$ Hamiltonian is modifyed as
\begin{equation}
\hat{W}'=\hat{W}+\frac{\alpha}{2}  \sum_x (-1)^{x}
[\hat{\chi}^\dagger 
(\mbox{\bf x}),\hat{\chi}(\mbox{\bf x})], 
\label{eq:W'}
\end{equation}
where $\alpha$ is an arbitrary parameter. 
Thus, $<\bar{\chi}\chi >$ is obtained in the standard way as
\begin{equation}
<\bar{\chi}\chi >= \frac{\partial \omega '}
{\partial \alpha} |_{\alpha =0}.
\end{equation}

The massless continuum Schwinger model 
undergoes a breaking of chiral symmetry with 
\begin{equation}
<\bar{\psi} \psi>/e = \frac{e^\gamma}{2\pi^{3/2}}=0.15995,
\label{eq:chiral-cont}
\end{equation}
where $\gamma$ is the Euler constant. 
This non-zero value of the chiral condensate is one of the main efects 
of the axial anomaly.

In FIG. 4 we report the value of the chiral condensate
per-lattice-site for
lattice sizes ranging from $N=2$ to $N=10$.
FIG. 5 shows this chiral order parameter for different lattice
sizes up to order $K=N-2$ for each size.

\begin{center}
\begin{figure}[t]
\hskip 1cm \psfig{figure=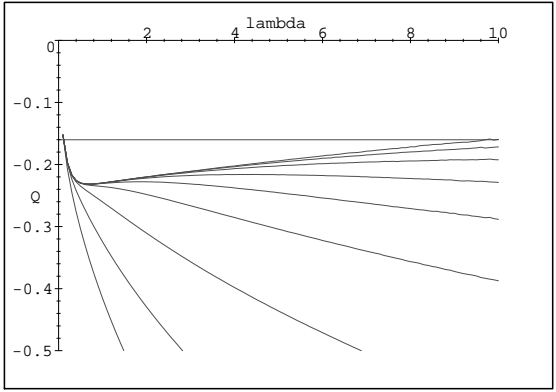 
,width=8.6cm}
\caption{The chiral condensate per-lattice-site 
times $\lambda^{1/2}$ vs. $\lambda$
for orders $K=1,2,\ldots,10$
on a lattice of size $N=10$ .
}
\label{fig4}
\end{figure}
\end{center}

\begin{center}
\begin{figure}[t]
\hskip 1cm \psfig{figure=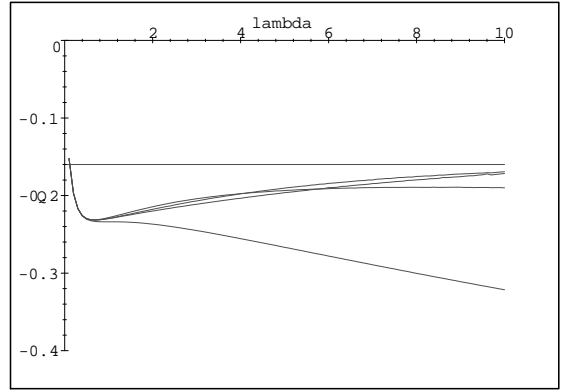,width=8.6cm}
\caption{The chiral condensate per-lattice-site
times $\lambda^{1/2}$ vs. $\lambda$
on lattices of size $N=4,6,8,10$ 
for order $K=N-2$ respectively.}
\label{fig5}
\end{figure}
\end{center}

Notice that the results 
in the weak coupling region converge to the corresponding continuum 
value (\ref{eq:chiral-cont}) 
as long as $N$
increases  while for a fixed $N$ the value improves with the
order $K$ till the value $K=N-2$ is reached.

\section{Conclusions and Final Remarks}

Our general proposal is to 
to show that the P-representation
is a valuable and alternative computational tool 
for gauge theories 
with dynamical fermions. In particular, in this work,
we wanted to test the Hamiltonian approach on finite lattices. 
With tis aim, we chose the simplest model: (1+1) QED.
This also enables us to compare with the corresponding numerical 
simulations \cite{f97} using the Lagrangian counterpart of the 
P-representation or the socalled 
{\em worldsheet formulation} \cite{afg}. This comparison shows that
, for this case of one spatial dimension, the Hamiltonian method is less
time consuming. 

The results are very good and confirm the belief of Hamer et al
\cite{hamer97} in obtaining with considerable 
accuracy the observables working on
lattices of moderate size. Consequently, this procedure is appealing 
because one can run all the needed computations in small machines 
obtaining quite fair results. 
 
Finally, we would like to stress (once more) 
that our aim was not to present another 
solution to the Schwinger model, but, to test an alternative general 
approach to tackle dynamical fermions.

\vspace{15mm}

{\large \bf Acknowledgements}

\end{document}